\documentclass[12pt,a4paper,aps,prd,superscriptaddress,nofootinbib,preprintnumbers]{revtex4-1}
\usepackage[utf8]{inputenc}
\usepackage{graphicx}
\usepackage{amssymb}
\usepackage{textcomp}
\usepackage{amsmath}
\usepackage{tabularx}
\usepackage{bm}
\usepackage{times}
\usepackage{color}
\usepackage{slashed}
\usepackage{multirow}
\usepackage{verbatim}
\usepackage{cancel}
\usepackage{subfigure}
\usepackage[normalem]{ulem}
\usepackage[sort&compress]{natbib}

\def\linkcolor{cyan!70!black}

\usepackage[colorlinks=true, pdfstartview=FitV, linkcolor=\linkcolor, citecolor=\linkcolor, urlcolor=\linkcolor]{hyperref}

\allowdisplaybreaks[4]

\usepackage{orcidlink}

\def\lsim{\mathrel{\raise.3ex\hbox{$<$\kern-.75em\lower1ex\hbox{$\sim$}}}}
\def\gsim{\mathrel{\raise.3ex\hbox{$>$\kern-.75em\lower1ex\hbox{$\sim$}}}}

\def\beq{\begin{equation}}
\def\eeq{\end{equation}}
\def\bea{\begin{eqnarray}}
\def\eea{\end{eqnarray}}
\def\nn{\nonumber}

\def \({\left(}
\def \){\right)}
\def \[{\left[}
\def \]{\right]}
\def \l|{\left|}
\def \r|{\right|}

\definecolor{orange}{rgb}{1,0.5,0}

\begin{document}

\title{\boldmath 
Is Dark Matter the origin of the \texorpdfstring{$B\to K \nu\bar\nu$}{B->Knunubar} excess at Belle~II?}

\author{Lorenzo Calibbi\,\orcidlink{0000-0002-9322-8076}}
\email{calibbi@nankai.edu.cn}
\author{Tong Li\,\orcidlink{0000-0002-6437-8542}}
\email{litong@nankai.edu.cn}
\author{Lopamudra Mukherjee\,\orcidlink{0000-0001-8765-7563}}
\email{lopamudra.physics@gmail.com}
\affiliation{
School of Physics, Nankai University, Tianjin 300071, China
}
\author{Michael A.~Schmidt\,\orcidlink{0000-0002-8792-5537}}
\email{m.schmidt@unsw.edu.au}
\affiliation{
Sydney Consortium for Particle Physics and Cosmology,\\
School of Physics, The University of New South Wales, Sydney, New South Wales 2052, Australia
}
\preprint{CPPC-2025-01}
\begin{abstract}
We present two models of dark matter~(DM) that can provide a natural explanation of the excess of \mbox{$B^+\to K^+ +\,\text{invisible}$} events with respect to the Standard Model~(SM) prediction for \mbox{$B^+\to K^+ \nu\bar\nu$}, which has been reported by the Belle~II collaboration. Interactions between the dark and the visible sector are mediated by an axion-like particle (ALP) in one case, by the kinetic mixing between a dark photon and the SM photon in the second case. Both models encompass a light fermion singlet as the DM candidate and can account for the observed DM relic abundance through, respectively, the freeze-in and the freeze-out production mechanism, while simultaneously explaining the Belle~II excess.
\end{abstract}

\maketitle

\section{Introduction}

Recently, the Belle II collaboration obtained the first evidence for $B^+\to K^+ \nu\bar\nu$~\cite{Belle-II:2023esi}, one of the rare $B$ meson decays, measuring 
\begin{align}
    \mathrm{BR}(B^+\to K^+ \nu\bar\nu)_{\rm Belle~II}=\left(2.3\pm 0.5\, \text{(stat)}^{+0.5}_{-0.4}\,\text{(syst)}\right)\times 10^{-5}\;.
\end{align}
In the Standard Model (SM), the decay is mediated by a $W$ boson exchange at the loop level and thus is suppressed. In addition, there is a tree-level long-distance contribution from $B^+ \to \nu_\tau \tau^+(\to K^+ \bar\nu_\tau)$, which accounts for about 10\% of the SM prediction and has been treated as a background in the experimental analysis. Together they result in~\cite{Parrott:2022zte}\footnote{See also Ref.~\cite{Becirevic:2023aov} that predicts a slightly smaller central value and Refs.~\cite{Altmannshofer:2009ma,Buras:2014fpa,Blake:2016olu} for other recent calculations of the branching ratio within the SM.}
\begin{align}
    \mathrm{BR}(B^+\to K^+\nu\bar\nu)_{\rm SM} = (5.58\pm0.37) \times 10^{-6}\,,
\end{align}
where the tree-level contribution accounts for $(0.61\pm0.06)\times 10^{-6}$, which implies a $2.7\sigma$ tension between the Belle~II measurement and the SM prediction. Combining the latest Belle II result with earlier measurements results in a weighted average of $\mathrm{BR}(B^+\to K^+\nu\bar\nu)_{\rm exp} = (1.3\pm0.4)\times 10^{-5}$~\cite{Belle-II:2023esi}.
An attractive explanation of the excess is an additional decay channel with undetected final states,\footnote{Explanations in terms of heavy new physics have been discussed in Refs.~\cite{Athron:2023hmz,Bause:2023mfe,Allwicher:2023xba,Chen:2023wpb,Chen:2024jlj,Hou:2024vyw,Marzocca:2024hua,Karmakar:2024gla,Hati:2024ppg,Allwicher:2024ncl,Tian:2024ubt,Alda:2024sup,Dev:2024tto,Becirevic:2024iyi,Buras:2024mnq,Bhattacharya:2024clv,Chen:2024cll,Kim:2024tsm,Zhang:2024hkn}.} such as sterile neutrinos~\cite{Browder:2021hbl,He:2021yoz,Felkl:2021uxi,Felkl:2023ayn,Ovchynnikov:2023von,Gartner:2024muk,Rosauro-Alcaraz:2024mvx,Buras:2024ewl,Becirevic:2024iyi,Datta:2023iln}, light dark matter (DM) particles~\cite{Bird:2004ts,Altmannshofer:2009ma,He:2022ljo,He:2023bnk,Hou:2024vyw,He:2024iju,Ho:2024cwk,Abdughani:2023dlr,Berezhnoy:2023rxx}, an axion-like particle (ALP)~\cite{Berezhiani:1989fs,Berezhiani:1990jj,Berezhiani:1990wn,Ferber:2022rsf,Altmannshofer:2023hkn,Altmannshofer:2024kxb,Dai:2024onu} or more generally long-lived particles, which escape the detector undetected~\cite{Filimonova:2019tuy,Fridell:2023ssf,Hu:2024mgf,Bolton:2024egx,Gabrielli:2024wys,McKeen:2023uzo,Davoudiasl:2024cee}.

Refs.~\cite{Altmannshofer:2023hkn,Fridell:2023ssf} pointed out that the excess is well described by a two-body decay $B^+\to K^+ X$, where the new bosonic particle $X$ has mass $m_X\simeq 2~\mathrm{GeV}$ and either escapes the detector undetected or decays into invisible final states. 
This scenario has been studied in the context of several explicit models~\cite{Abdughani:2023dlr,Berezhnoy:2023rxx,McKeen:2023uzo,Davoudiasl:2024cee,Hu:2024mgf,Bolton:2024egx,Altmannshofer:2024kxb,Dai:2024onu}. Collider experiments set the most stringent constraints on a GeV-scale bosonic particle. Relevant searches include the LHCb displaced vertex search for $B\to K^{(*)} X (\to\mu\bar\mu)$~\cite{LHCb:2015nkv,LHCb:2016awg}, the corresponding prompt decay search~\cite{LHCb:2014cxe}, BaBar's search for dark photons in $e^+ e^- \to \gamma X$~\cite{BaBar:2014zli,BaBar:2013npw}, and a reanalysis~\cite{Dobrich:2018jyi} of CHARM data~\cite{CHARM:1985anb}. For sub-GeV bosonic particles, there are additional constraints from astrophysical observations~\cite{Raffelt:1990yz} and beam dump experiments~\cite{PBC:2025sny}, which however currently do not extend to $m_X\simeq 2$ GeV, because the lower centre-of-mass energy in the existing beam dump experiments and the low temperature of astrophysical objects results in an exponential suppression of the production of bosonic particles with $m_X\simeq2$ GeV mass. 

In this work, we focus on the connection of the excess in $B^+\to K^+ +\mathrm{inv}$ to light sub-GeV dark matter. An important requirement for all DM models is the explanation of the observed cosmic DM relic abundance of $\Omega_{\rm DM} h^2 = 0.12$~\cite{Planck:2018vyg}, which has been previously discussed within three models~\cite{He:2024iju,Ho:2024cwk,Abdughani:2023dlr}.\footnote{
See Refs.~\cite{Bharucha:2022lty,Ghosh:2023tyz, Cheek:2022yof, Babu:2023zni, Mandal:2023jnv} for earlier work on freeze-in DM with an ALP mediator and Refs.~\cite{Pospelov:2007mp,Frandsen:2011cg,Feng:2017drg,Knapen:2017xzo,Dutra:2018gmv,Jung:2020ukk,Rizzo:2020jsm,Borah:2020swo,Rizzo:2024bhn,Garcia:2024uwf,Krnjaic:2025noj,Alonso-Gonzalez:2025xqg} for freeze-out DM models with a GeV-scale dark photon mediator. 


} 
In Ref.~\cite{He:2024iju} a scalar DM candidate produced via the freeze-out mechanism is introduced, with the observed relic density achieved through pair annihilation of dark matter to light mesons. The excess in $B^+\to K^+ +\mathrm{inv}$ is explained via heavy vector-like quark mediators which couple the scalar DM to the bottom and strange quark. The scenario is already being probed by direct detection experiments through the Migdal effect and requires mild tuning. 

Ref.~\cite{Ho:2024cwk} also considers scalar DM produced via the freeze-out mechanism, but it relies on a light scalar mediator to address the excess in $B^+\to K^+ +\mathrm{inv}$. The authors explain it by a two-body decay into the scalar mediator and a three-body decay for a slightly heavier scalar mediator with a mass larger than $m_{B^+}-m_{K^+}$. In both cases the scalar mediator dominantly decays to light sub-GeV DM. DM direct detection has not been discussed in detail for this scenario. We expect that the Migdal effect will further constrain the model and possibly rule out the explanation of the excess in $B^+\to K^++\mathrm{inv}$.

Finally, Ref.~\cite{Abdughani:2023dlr} considers a scalar mediator $\varphi$ together with the fermionic DM field $\chi$. The explanation of the observed DM relic abundance requires resonant DM annihilation, and thus $m_\varphi \simeq 2m_\chi$. The excess in $B^+\to K^+ +\mathrm{inv}$ is accounted for by the decay $B^+\to K^+ \varphi(\to \chi\bar\chi)$. However, a large part of the parameter space is strongly constrained by LHCb searches~\cite{LHCb:2015nkv,LHCb:2016awg} for long-lived scalar particles using $B\to K^{(*)}\mu\bar\mu$, whose complementarity to $B\to K^{(*)} + \mathrm{inv}$ has been discussed in detail in Ref.~\cite{Ovchynnikov:2023von}.

We revisit the fermionic DM scenario with a bosonic mediator and demonstrate the possibility of explaining both the excess in $B^+\to K^++\mathrm{inv}$ and the observed DM relic abundance, while naturally avoiding all other constraints without relying on fine-tuning of the models' parameters. Two well-motivated mediators are an axion-like particle (ALP) and a dark photon. In Section~\ref{sec:ALP}, we discuss the ALP portal model, while Section~\ref{sec:DP} presents the phenomenology of a dark-photon portal model. Finally, we conclude in Section~\ref{sec:Conclusions} with a comparison of the two scenarios and an outlook.

\section{ALP portal model}
\label{sec:ALP}

We start considering the most general effective Lagrangian for the interactions of an ALP with the SM fermions ($f=u,d,\ell$, respectively denoting  up-type quarks, down-type quarks, charged leptons\,\footnote{We neglect ALP interactions with neutrinos, since they are suppressed by the tiny mass of the latter.}) that is given by~\cite{Brivio:2017ije,Bauer:2021mvw}
\begin{eqnarray}
\mathcal{L}_{aff}
&=& \frac{\partial_\mu a}{2f_a} \bar{f}_i \gamma^\mu \left(C^V_{f_i f_j}+C^A_{f_i f_j}\gamma_5\right) f_j\;,
\label{eq:Lff}
\end{eqnarray}
where the ALP decay constant $f_a$ is an energy scale related to the spontaneous breaking of an underlying global ${U}(1)$ symmetry, $i,j=1,2,3$ are flavour indices, and the dimensionless couplings are such that $(C^{V,A}_{f_i f_j})^\ast=C^{V,A}_{f_j f_i}$ for $i\neq j$, while the vector-current interactions $(C^{V}_{f_i f_j})$ are unphysical for $i=j$.

In addition, even if not present in the UV theory (e.g.~because ${U}(1)$ is anomaly free), ALP couplings to photons and gluons are induced by fermion loops:
\begin{eqnarray}
\mathcal{L}_{G}
= 
C^\text{eff}_\gamma\,\frac{\alpha}{4\pi}\frac{a}{f_a} F_{\mu\nu} \tilde{F}^{\mu\nu} + 
C^\text{eff}_G\,\frac{\alpha_s}{4\pi}\frac{a}{f_a} G^a_{\mu\nu} \tilde{G}^{\mu\nu,\,a}
\;,
\label{eq:Lgauge}
\end{eqnarray}
where we use the convention $\tilde{X}^{\mu\nu}\equiv \frac12\varepsilon^{\mu\nu\alpha\beta}X_{\alpha\beta}$ with $\varepsilon^{0123}=1$. 
As mentioned, these effective couplings are in general given by a model-dependent UV coupling plus the contribution of SM fermion loops~\cite{Bauer:2021mvw}:
\begin{equation}
    C_{\gamma}^{\rm eff} = C^\text{UV}_{\gamma} + \sum_{f} C^A_{ff} N_c^f Q_f^2 \, B\left(\frac{4 m_f^2}{m_a^2}\right)\;,\quad C_{G}^{\rm eff} = C^\text{UV}_{G} + \frac12 \sum_{q} C^A_{qq}\, B\left(\frac{4 m_q^2}{m_a^2}\right)\;,
    \label{eq:Cgauge}
\end{equation}
where the first sum runs over the flavour-conserving ALP couplings to all SM fermions (whose electric charge is $Q_f$ and number of colours is $N_c^f$), while the second sum only runs over (up and down) quarks.
The loop function
\begin{equation}
    B(x) = 1-x f^2(x), 
    \qquad 
    f(x) = \begin{cases} \arcsin \tfrac{1}{\sqrt{x}} & x \geq 1 \\
    \tfrac\pi2+\tfrac{i}{2}\ln\frac{1+\sqrt{1-x}}{1-\sqrt{1-x}} & x<1
    \end{cases}\;,
\end{equation}
is such that the contribution of a given fermion vanishes in the decoupling limit $m_a \ll m_f$.

Finally, we introduce an ALP-DM interaction, analogous to the fermion interactions in Eq.~\eqref{eq:Lff}, involving a singlet Dirac fermion $\chi$ that plays the role of our DM candidate:
\begin{eqnarray}
\mathcal{L}_{a\chi\chi}
&=& \frac{\partial_\mu a}{2f_a} \bar{\chi} C^A_{\chi\chi}\gamma^\mu \gamma_5 \chi\;.
\label{eq:Lchi}
\end{eqnarray}

\mathversion{bold}
\subsection{Contribution to \texorpdfstring{$B\to K^{(*)} + $ invisible}{B->K(*)+invisible}}
\mathversion{normal}
\label{sec:ALPBKnunu}
In the presence of flavour-violating interactions with $b$ and $s$ quarks and if the ALP is sufficiently light, the above Lagrangian will induce two-body decays of $B$ mesons to a kaon and an ALP. 
The resulting branching ratios read~\cite{Altmannshofer:2023hkn}
\begin{eqnarray}
\label{eq:BtoK}
\mathrm{BR}(B \to K a) &=& \frac{\tau_B m_B^3}{64\pi}\frac{|C_{sb}^V|^2}{f_a^2}\(1-\frac{m_K^2}{m_B^2}\)^2 f_0^2(m_a^2)\lambda^{1/2}\left(1,\frac{m_K^2}{m_B^2},\frac{m_a^2}{m_B^2}\right)\;,\\
\mathrm{BR}(B \to K^{\ast} a) &=& 
\frac{\tau_B m_B^3}{64\pi}\frac{|C_{sb}^A|^2}{f_a^2}A_0^2(m_a^2)\lambda^{3/2}\left(1,\frac{m_K^2}{m_B^2},\frac{m_a^2}{m_B^2}\right)\;,
\label{eq:BtoK*}
\end{eqnarray}
where $f_0(q^2)$ and $A_0(q^2)$ are the relevant $B \to K^{(*)}$ scalar form factors~\cite{Gubernari:2023puw} (see also \cite{Gubernari:2018wyi}) to be evaluated at $q^2\equiv m_a^2$ and $\lambda(x,y,z) \equiv x^2 + y^2 + z^2 - 2(xy + yz + zx)$ denotes the K\"all\'en function. Note that, while the decay to the pseudoscalar kaon depends on the vector couplings $C^V_{sb}$, the decay to the vector meson $K^*$ is controlled by the axial-vector coupling $C^A_{sb}$ and hence the two decays are sensitive to different parameters. 

\begin{figure}[t!]
    \centering
    \includegraphics[width=0.5\textwidth]{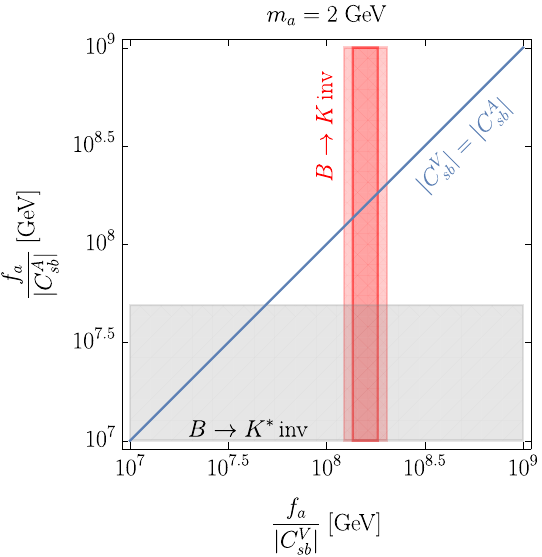}
    \caption{ Illustration of the
    parameter space of the ALP model that is able to explain the observed Belle~II excess in $B\to K+ \mathrm{inv}$ at the $1\sigma$ (dark red band) and $2\sigma$ level (light red band) for $m_a = 2~\mathrm{GeV}$. The gray region is excluded at $95\%$ CL by searches for $B\to K^{*} + {\rm inv}$. See the main text for details.
    }
    \label{fig:alp-flavour}
\end{figure}

If the ALP is long-lived or decays invisibly, the two-body decay $B \to K a$ can provide a good fit to the $\mathrm{BR}(B \to K + {\rm inv}$) measured by Belle~II~\cite{Altmannshofer:2023hkn,Fridell:2023ssf}.\footnote{If instead $m_a > m_B$, hence the ALP is off-shell, the $q^2$ distribution of the three-body processes $B\to K a^{(*)} \to K \chi\bar\chi$ does not give a good fit to the Belle~II $B\to K + {\rm inv}$ data, because the mediator is a (pseudo)scalar field~\cite{Fridell:2023ssf}.} 
The Belle~II excess requires $m_a \approx 2~\mathrm{GeV}$ and $\mathrm{BR}(B \to K a) = (0.5-0.9)\times 10^{-5}$ at 1$\sigma$, $(0.4-1.1)\times 10^{-5}$ at 2$\sigma$~\cite{Fridell:2023ssf}. Using Eq.~\eqref{eq:BtoK}, we find that
this translates into the following range for the vector coupling of the ALP to $b$ and $s$: 
\begin{equation}
1.4 \times 10^8 ~\mathrm{GeV} < f_a/|C^V_{sb}| < 1.8 \times 10^8~\mathrm{GeV}\quad (1\sigma)\,,    
\end{equation}
which is shown as red region in Fig.~\ref{fig:alp-flavour}.

As we have seen above, $\mathrm{BR}(B \to K^{\ast} a)$ is instead sensitive to the axial-vector coupling $C^A_{sb}$. We find that, for an ALP with $m_a=2~\mathrm{GeV}$, this process is at odds with the BaBar search for \mbox{$B\to K^{*}\nu\bar\nu$}~\cite{Lees:2013kla} unless $f_a/|C^A_{sb}|> 4.9\times 10^7~\mathrm{GeV}$, i.e.~the gray shaded region in Fig.~\ref{fig:alp-flavour} is excluded.
Hence, natural scenarios with $|C^A_{sb}| = |C^V_{sb}|$\,---\,following, for instance, from an ALP coupling only to left-handed ($C^A_{sb} = - C^V_{sb}$) or right-handed ($C^A_{sb} = C^V_{sb}$) quarks\,---\,are compatible with both the Belle~II excess and the BaBar constraint on 
$B\to K^{*} + {\rm inv}$ if we set \mbox{$f_a/|C^A_{sb}| = f_a/|C^V_{sb}| \approx 10^{8}~\mathrm{GeV}$}.\footnote{However, it should be noted that the BaBar $B\to K + {\rm inv}$ data in Ref.~\cite{Lees:2013kla} are in mild tension with the Belle~II measurement~\cite{Belle-II:2023esi}, which leads to a deterioration of the $\chi^2$ of all new physics hypotheses considered in Ref.~\cite{Fridell:2023ssf}.}
Invisible signals require the ALP to escape the detector or decay inside the detector into dark matter and thus also escape detection, which depends on the lifetime of the ALP, its mass and the hierarchy of its couplings.
Thus, the NP contribution to the invisible branching ratios we are interested in receives two contributions:
\begin{align}\label{eq:BRinv}
    {\rm BR}(B\to K^{(*)} + {\rm inv})_{\rm NP}={\rm BR}(B\to K^{(*)} a) \left[ {\rm BR}(a\to {\rm \chi\bar\chi}) + e^{-L_{\rm det}/\beta\gamma c \tau_a} \mathrm{BR}(a\to {\rm vis})\right]\;,
\end{align}
where $c\tau_a$ is the proper decay length of the ALP, $\beta\gamma$ is the ALP velocity times its boost factor, and $L_{\rm det}$ denotes the typical detector size, which is $L_{\rm det} \approx 3.5~\mathrm{m}$ for Belle~II~\cite{Kou:2018nap}. The quantity \mbox{$\mathrm{BR}(a\to {\rm vis}) = \mathrm{BR}(a\to f\bar f,gg,\gamma\gamma)$} is the total branching ratio of the visible ALP decays into fermion-antifermion pairs, gluons, two photons. We do not include the contribution of visible ALP decays close to the production point near the beam pipe but outside the detector coverage. Although these decays are relevant for short-lived ALPs~\cite{Ferber:2022rsf}, their contribution is subdominant for the coupling regime we consider here, which leads to a long-lived ALP. See the discussion below.

The ALP decay width into Dirac DM particles is given by\,\footnote{For Majorana DM the branching ratio has to be multiplied by a factor of 2 due to the factor 2 larger matrix element and the symmetry factor $1/2$.}
\begin{equation}
\Gamma(a\to \chi \bar{\chi}) = \frac{|C_{\chi\chi}^A|^2 m_a m_\chi^2}{8\pi f_a^2}  \sqrt{1-\frac{4m_\chi^2}{m_a^2}}\;.
\label{eq:ALPtoDM}
\end{equation}
An analogous expression holds for the (flavour-conserving) decays into the kinematically accessible SM fermions:
\begin{equation}
\Gamma(a\to f \bar{f}) = \frac{|C_{ff}^A|^2 N_c^f  m_a m_f^2}{8\pi f_a^2}  \sqrt{1-\frac{4m_f^2}{m_a^2}}\;,
\end{equation}
where one can see that the ALP preferably decays into the heaviest available final state.
We omit possible flavour-violating decay modes that are irrelevant for the following discussion. Similarly, for simplicity, we neglect the leptonic decays without qualitatively affecting our results.

For $m_a = 2~\mathrm{GeV}$, the above perturbative expression provides a good approximation for the decays into quarks. However, the hadronic width of the ALP is to a large extent dominated by decays to gluon pairs~\cite{Bauer:2021mvw,DallaValleGarcia:2023xhh}, whose width reads
\begin{align}
\Gamma(a\to gg) = |C_G^{\rm eff}|^2 \,\frac{\alpha_s^2m_a^3}{8\pi^3 f_a^2}\left(1+\frac{83\alpha_s}{4\pi}\right)\;.
\end{align}
Finally, the ALP decay width into photons is given by:
\begin{align}
\Gamma(a\to \gamma\gamma) = |C_\gamma^{\rm eff}|^2 \,\frac{\alpha^2m_a^3}{64\pi^3 f_a^2}\;.
\end{align}
For the following values of the parameters
\begin{equation}
m_a=2~\mathrm{GeV}\,,\quad C_G^\text{UV}=C_\gamma^\text{UV}=0\,,\quad f_a/C_q=10^8~\mathrm{GeV}\,,
\label{eq:ALPparam}
\end{equation}
where $C_q$ is a universal coupling to quarks, $C^{A}_{qq} = C^{A,V}_{sb} \equiv C_q$, and using $\alpha_s(m_a) \approx 0.3$~\cite{ParticleDataGroup:2024cfk}, the above expressions give for the visible decay modes:
\begin{align}
    \mathrm{BR}(a\to gg) \simeq 93\%\,,\quad
    \mathrm{BR}(a\to s\bar s) \simeq 7\%\,, \quad
    \mathrm{BR}(a\to \gamma\gamma) \simeq 2\times 10^{-5}
\end{align}
for negligible decays to dark matter. This is consistent with the results in Ref.~\cite{DallaValleGarcia:2023xhh}. The decay into DM is subdominant for $m_\chi \ll m_a$:
\begin{align}
 \mathrm{BR}(a\to \chi\bar\chi) \simeq 2.6\times 10^{-6}\times\left(\frac{C^A_{\chi\chi}}{C_q}\right)^2 \left(\frac{m_\chi}{1~\mathrm{MeV}}\right)^2\, .
\end{align}
The resulting ALP decay length is
\begin{align}\label{eq:ctau}
 c\tau_a \simeq 65\,\mathrm{m}\times\left(\frac{10^{8}~\mathrm{GeV}}{f_a/C_q}\right)^2 \,.
\end{align}
Hence, the ALP decays visibly but is long-lived. From Eqs.~\eqref{eq:BRinv} and \eqref{eq:BtoK}, it follows that 
\begin{equation}
{\rm BR}(B^+\to K^+ + {\rm inv})_{\rm NP} \simeq {\rm BR}(B^+\to K^+ a) \simeq 1.7 \times 10^{-5} \times \left(\frac{10^8~\mathrm{GeV}}{f_a/C^V_{sb}}\right)^2 \,,    
\end{equation}
and, thus, the Belle~II excess can be indeed accommodated by a natural choice of the parameters as the one in Eq.~\eqref{eq:ALPparam}.

\subsection{Dark Matter freeze-in production}

Even assuming a vanishing ALP abundance in the early universe, the flavour-conserving and flavour-violating interactions considered above will produce an ALP population through $b$ quark decay and scattering processes, as discussed in Ref.~\cite{Aghaie:2024jkj}. Here, in contrast to that work, the ALP is not itself the DM candidate but it decays into SM particles and, subdominantly, to DM particles, creating a DM population via the freeze-in mechanism~\cite{McDonald:2001vt,Hall:2009bx}. Furthermore, Ref.~\cite{Aghaie:2024jkj} considered ALP freeze-in production while, in our case, the ALP is quickly thermalised by processes involving $b$ quarks, as we will show in the following.  
The relevant processes for ALP production are the decays
\begin{align}
    \Gamma(b\to s a) = \frac{m_b^3}{64\pi \,f_a^2} \left( |C^V_{sb}|^2 + |C^A_{sb}|^2\right) \left(1-\frac{m_a^2}{m_b^2}\right)^3\,,
\end{align}
and the following scattering processes, whose cross-sections, reinstating the ALP mass dependence in the formulae of Ref.~\cite{Aghaie:2024jkj}, reads
\begin{align}
    \sigma(b\bar b \to a g)  =&~ \frac{4\alpha_s |C^A_{bb}|^2}{9 f_a^2} \frac{x_b}{(1-4x_b)(1-x_a)} \nn\\
    & \times \left( (1+x_a^2-4 x_a x_b) \tanh^{-1}(\sqrt{1-4x_b}) -x_a \sqrt{1-4x_b} \right)\;,
    \\
    \sigma(b g \to ba) = &~ \frac{\alpha_s |C^A_{bb}|^2}{48 f_a^2} \frac{x_b}{(1-x_b)^3} \Bigg[
    \lambda_{ba}^{1/2} \left(7x_a+2x_ax_b -x_a x_b^2-\left(3-x_b\right) \left(1-x_b\right)^2\right) \nn \\ 
    & +4 \left(\lambda_{ba} +x_a^2\right) \coth ^{-1}\left(\frac{1+x_b-x_a}{\lambda_{ba}^{1/2}}\right)
    \Bigg]\;,
\end{align}
where $x_a \equiv m_a^2/s$, $x_b \equiv m_b^2/s$, with $s$ denoting the centre of mass energy squared of the collision, and $\lambda_{ba}\equiv\lambda(1,x_a,x_b)$. The cross sections for the analogous processes involving photons, $b\bar b\to a\gamma$ and $b\gamma\to ba$, are obtained by replacing, respectively, $4\alpha_s/9 \to Q_b^2\alpha/3$ and $\alpha_s/6\to Q_b^2\alpha$ in the above expressions.

In order to study the thermalization of the ALP, we calculate the thermal averages of the above processes. As customary, the thermally averaged decay width is
\beq
\langle \Gamma(b\to s a)\rangle = \Gamma(b\to s a) \frac{K_1(m_b/T)}{K_2(m_b/T)}\;,
\eeq
where $T$ is the temperature of the thermal bath, and $K_1$ and $K_2$ are modified Bessel functions of the first and second type respectively.
The averaged cross sections are given by:
\beq
\langle\sigma_{12\to 34} v\rangle = \frac{1}{8T \prod_{i=1}^2 m_i^2K_2(m_i/T) } \int_{(m_1+m_2)^2}^\infty ds \frac{\lambda(s,m_1^2,m_2^2)}{\sqrt{s}}K_1\({\sqrt{s}}/{T}\)\sigma_{12\to 34}\;,
\eeq 
where $s$ is the squared centre-of-mass energy of the collision and, for massless initial state particles, we use the $m_i\to 0$ limit value $m_i^2 K_2(m_i/T) = 2T^2$.
These quantities are to be compared with the Hubble parameter,
\beq 
H(T) = \frac{\pi}{3 M_\text{Pl}}\sqrt{\frac{g_{\star  \rho}(T)}{10}}\,T^2\,,
\eeq 
where $M_\text{Pl} = 2.4 \times 10^{18}~\mathrm{GeV}$, is the reduced Planck mass and $g_{\star \rho}$ is the total number of relativistic degrees of freedom present in the thermal bath at temperature $T$.

\begin{figure}[t!]
    \centering
    \includegraphics[width=0.41\textwidth]{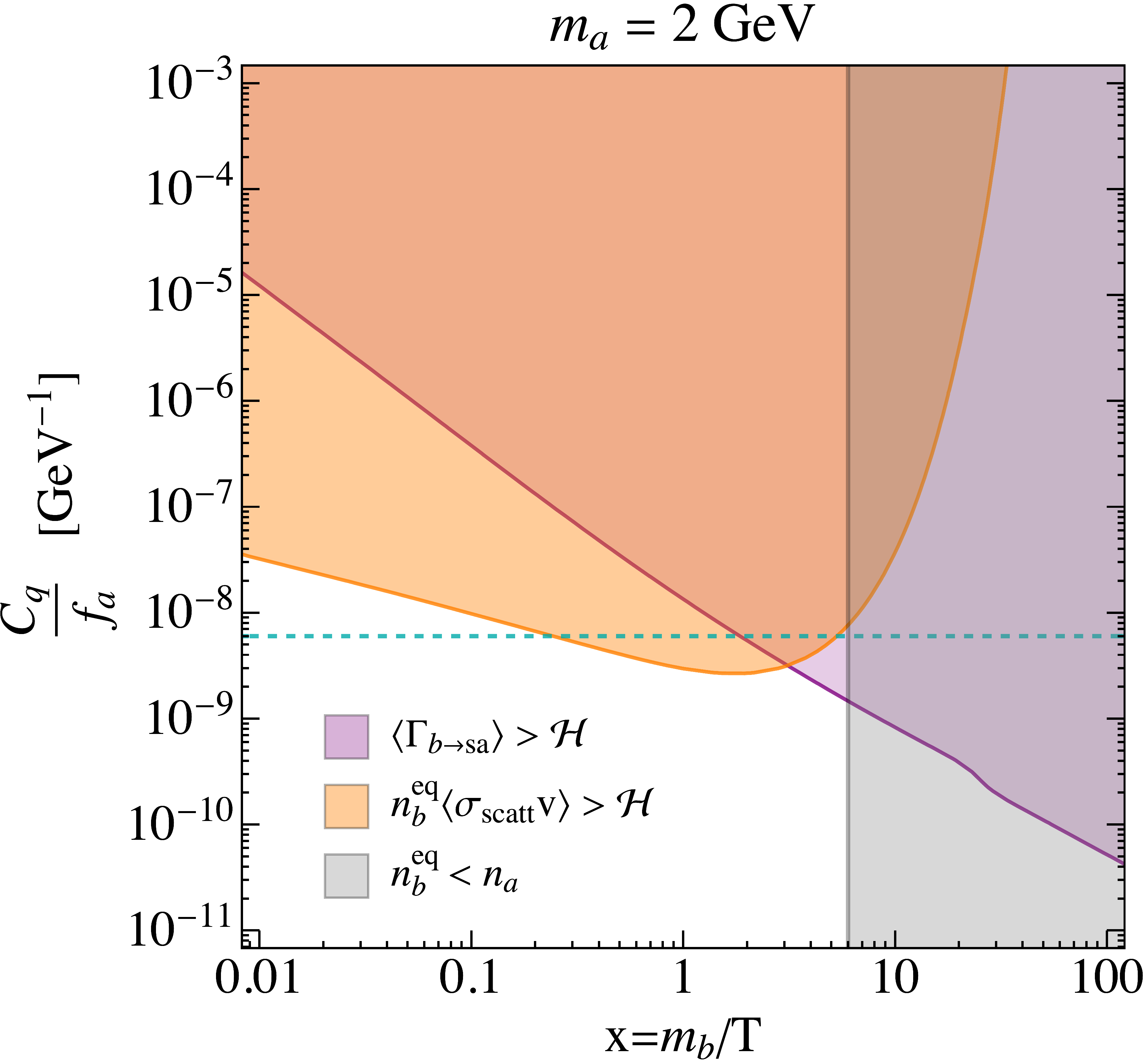}
    \hfill
    \includegraphics[width=0.58\textwidth]{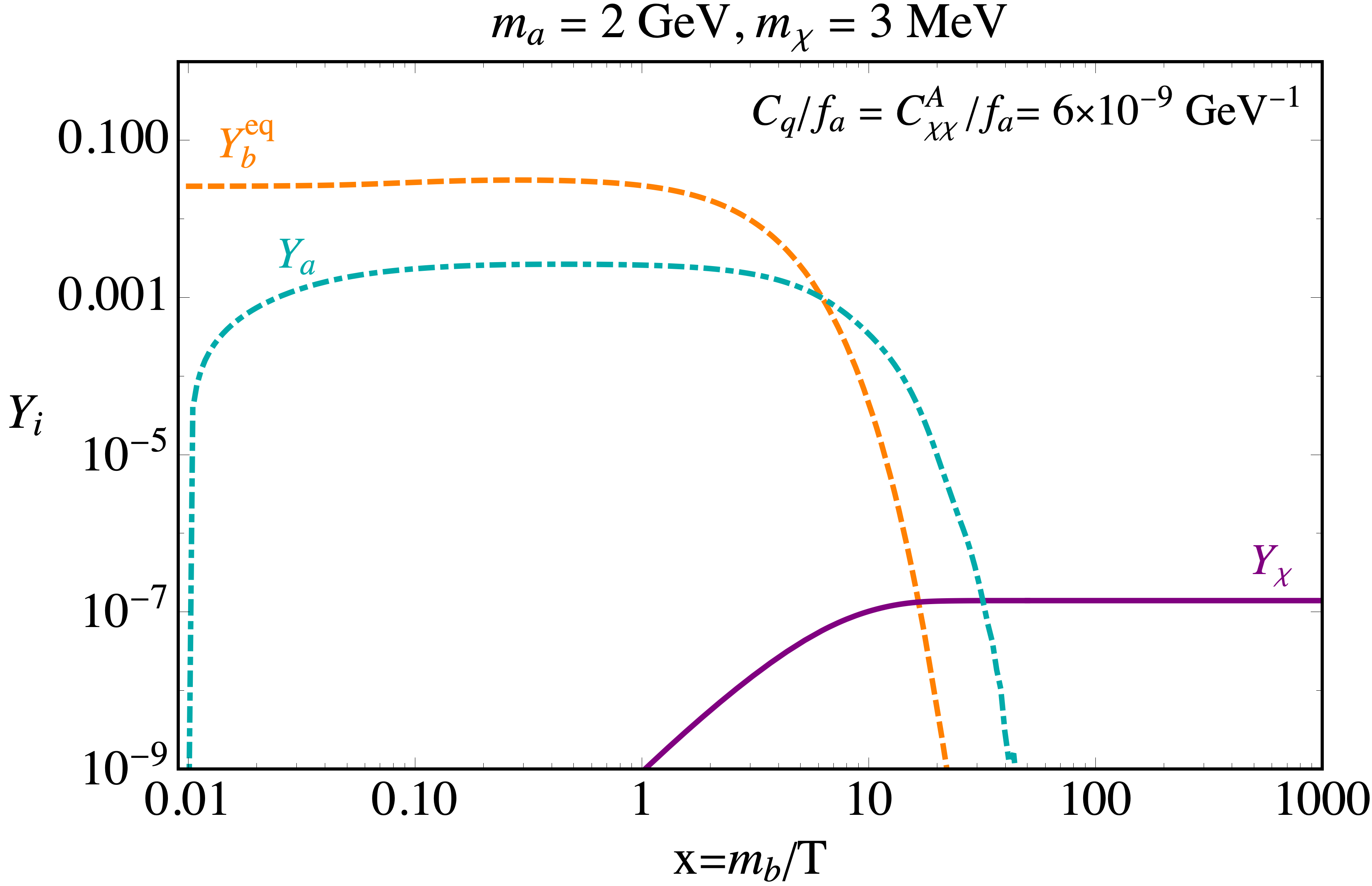}
    \caption{\emph{Left}: The shaded area shows the region where $b$ decays (purple) or scattering processes (orange) keep a 2~GeV ALP in equilibrium with the thermal bath. The gray region shows the region where the number density of $b$-quarks is too suppressed to keep the ALP in thermal equilibrium. The dashed blue line indicates the central value of the axion coupling $C_q/f_a$ required to explain the excess observed by Belle II in $B^+\to K^+ +\mathrm{inv}$.
    \emph{Right}: Evolution of different particle yields as a function of $x = m_b/T$ for universal ALP couplings $C_{q}/f_a = C_{\chi\chi}^A/f_a = 6\times 10^{-9}~\rm{GeV}^{-1}$, and ALP mass of $m_a = 2~\mathrm{GeV}$ and DM mass of $m_\chi = 3~\mathrm{MeV}$ and a vanishing UV coupling to gluons $ C_G^\text{UV}=0$. The dot-dashed cyan curve represents the ALP yield while the solid purple curve shows the dark matter yield. The dashed orange curve shows the equilibrium yield of the $b$ quark.
    }
    \label{fig:yield-alp}
\end{figure}

The left panel of Fig.~\ref{fig:yield-alp} shows that the above processes (and the corresponding inverse ones) quickly bring the ALP in thermal equilibrium with the SM bath. For a given value of the ALP coupling to quarks, the orange and purple regions, respectively, denote the temperature range for which the thermally averaged decay width and (normalised) scattering cross sections become larger than the Hubble parameter, thus corresponding to rates fast enough to overcome the expansion of the universe. In particular, for values of the coupling compatible with the Belle~II excess, $f_a/ C_q \approx 10^8~\mathrm{GeV}$, the scattering processes thermalise the ALP for $T \approx 10\times m_b$.\footnote{This estimate neglected flavour-violating scattering processes, such as $b\,s\to a\,g$, that are discussed in Ref.~~\cite{Aghaie:2024jkj}. In our scenario, their only effect would be to thermalise the ALP (slightly) earlier and this would not affect the final DM abundance and our following results.} 

The above conclusion is confirmed by the right panel of Fig.~\ref{fig:yield-alp}. Here, for representative values of the parameters, we show a numerical solution of the coupled Boltzmann equations for the ALP and the DM yields in terms of a time variable defined as $x \equiv m_b /T$:
\begin{align} 
\frac{dY_a}{d\ln x} H =& ~ Y^{eq}_b\langle\Gamma_{b\to sa}\rangle\(1-\frac{Y_a}{Y_a^{\rm eq}}\) + \nn \\
& s Y_b^{\rm eq} Y_{\gamma/g}^{\rm eq} \langle\sigma_{b\gamma/g \to b a} v\rangle\(1 - \frac{Y_a}{Y_a^{\rm eq}}\) + \nn \\
& s \left(Y_{b}^{\rm eq}\right)^2 \langle\sigma_{b \bar b \to \gamma/g a} v\rangle\(1 - \frac{Y_a}{Y_a^{\rm eq}}\) + \nn \\
& Y_a^{\rm eq}\langle \Gamma_{a\to \mathrm{SM}}\rangle \left(1-\frac{Y_a}{Y_a^{\rm eq}}\right) +
Y_a^{\rm eq} \langle \Gamma_{a\to \chi\bar\chi}\rangle \left[\left(\frac{\tilde Y_\chi}{\tilde Y_\chi^{\rm eq}}\right)^2 -\frac{Y_a}{Y_a^{\rm eq}}\right]
\;,
\\
\frac{d \tilde Y_\chi}{d\ln x} H = &  ~ 2 \,Y_a^{\rm eq}\langle\Gamma_{a\to \chi \bar \chi}\rangle \left[ \frac{Y_a}{Y_a^{\rm eq}}
-\left(\frac{\tilde Y_\chi}{\tilde Y_\chi^{\rm eq}}\right)^2 \right]\;,
\end{align}
where $\tilde Y_\chi \equiv Y_\chi + Y_{\bar\chi}$, equilibrium distributions are assumed for SM particles and the thermally averaged rates are normalized to the number densities of the indicated initial state. Furthermore, $a\to \text{SM}$ refers to the ALP decays to the SM states discussed above. The equilibrium yield of a given particle $i$ is given by
\begin{equation}\label{eq:eq-yield} 
Y_i^{eq} = \frac{n^{eq}_i}{s} = \frac{45}{4\pi^4}\frac{g_i}{g_{\star s}}\(\frac{m_i}{T}\)^2 K_2(m_i/T),
\end{equation}
where $s$ now denotes the entropy density, $g_{\star s}$ is the number of degrees of freedom contributing to it, and $g_i$ is the number of intrinsic degrees of freedom of $i$. In the numerical solution, we use the analytic approximation for $g_{\star s}$ from App.~A in Ref.~\cite{Wantz:2009it}. 

The right plot of Fig.~\ref{fig:yield-alp} illustrates that, for values of its couplings with quarks compatible with the Belle~II excess, the ALP is quickly produced and thermalised in the early universe. Subsequently, DM production through the process $a\to\chi\bar\chi$ takes place and is mainly effective at low temperatures just before the ALP density undergoes Botzmann suppression, as typical of the freeze-in mechanism through decays~\cite{Hall:2009bx}. The resulting DM yield $\tilde Y^{0}_\chi \equiv \tilde Y_\chi(T\ll m_a)$ can be employed to calculate today universe DM abundance in the usual way. For the numerical example of the figure, the resulting relic density is of the order of the value inferred by CMB observations, $\Omega_{\rm DM} h^{2} = 0.12$~\cite{Planck:2018vyg}:
\begin{align}
\Omega_\chi h^2 = \frac{m_\chi \tilde Y^0_\chi s_0}{3 H_0^2 M_\text{Pl}^2/h^2}  \approx 0.1\times
\left(\frac{m_\chi}{3\,{\rm MeV}} \right)^3
\left(\frac{|C^A_{\chi\chi}|/f_a}{6\times 10^{-9}~{\rm GeV}^{-1}} \right)^2 \,,
\end{align}
where we used the present universe values of the Hubble parameter and entropy density that, in natural units, are respectively $H_0 \simeq 2.13\, h\times 10^{-42}~\mathrm{GeV}$, $s_0 \simeq 2.22 \times 10^{-38}~\mathrm{GeV}^3$.
The parametric dependence of the relic density directly follows from the width of the production process, Eq.~(\ref{eq:ALPtoDM}).

\begin{figure}[t!]
    \centering
    \includegraphics[width=0.8\textwidth]{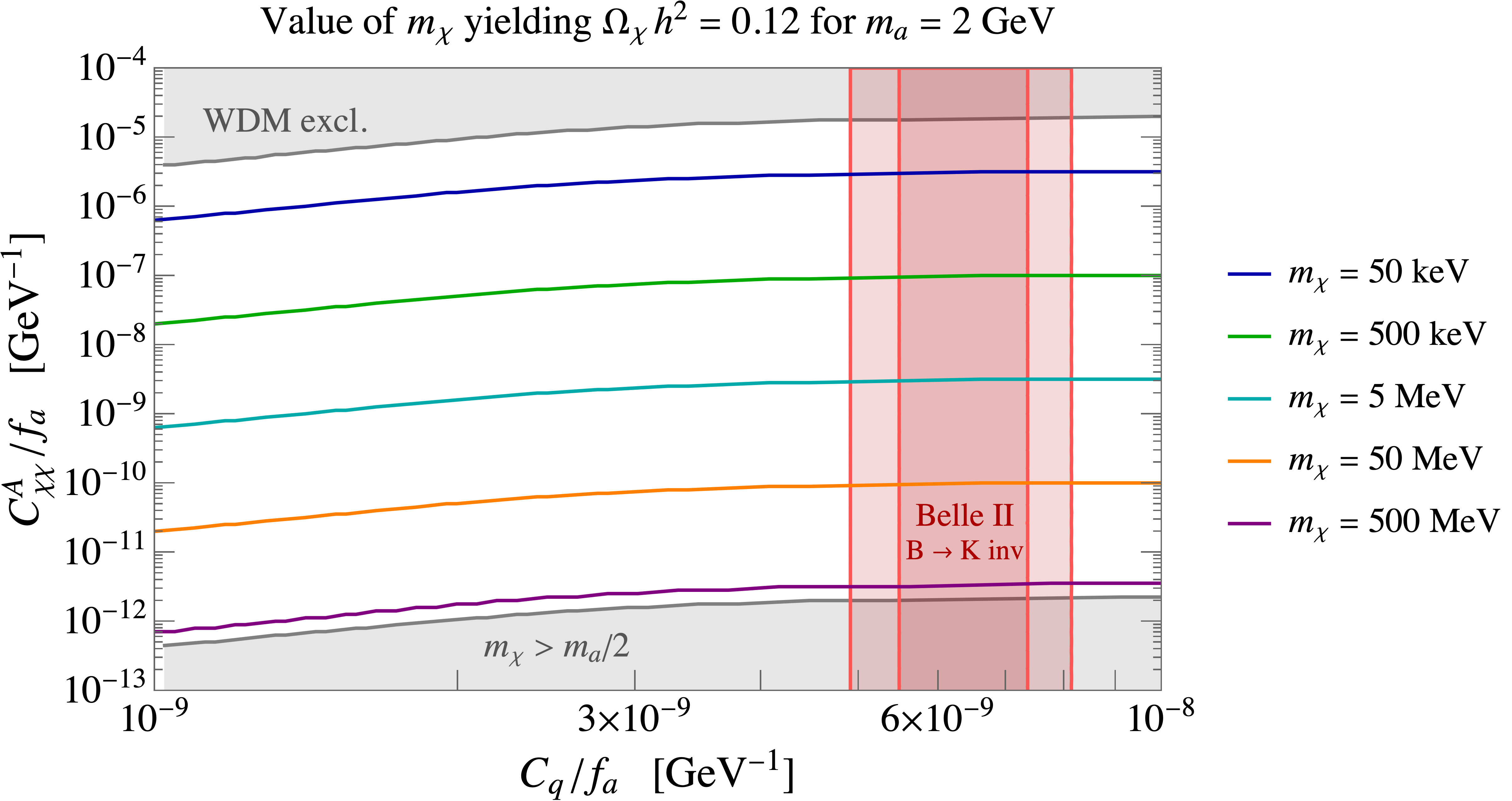}
    \caption{Contours of the DM mass $m_\chi$ corresponding to the observed relic abundance $\Omega_\chi h^2 = 0.12$ on the ($C^{V,A}_{bs}/f_a = C_{qq}^A/f_a \equiv C_q/f_a$, $C^A_{\chi\chi}/f_a$) plane. The $1\sigma$ ($2\sigma$) range of $C_q/f_a$ preferred by $B^+\to K^++{\rm inv}$ data is shown as a dark (light) red band. In the upper gray region, a DM abundance not exceeding the observed amount can only be obtained for values of the DM mass $m_\chi \lesssim 15~\mathrm{keV}$, which are in conflict with structure formation bounds on warm DM~(WDM)~\cite{Ballesteros:2020adh,DEramo:2020gpr,Decant:2021mhj}. In the lower region DM production through the decay $a\to \chi\bar\chi$ is kinematically forbidden.}
    \label{fig:ALP-summary}
\end{figure}

\subsection{Discussion}
The above results are summarized in Fig.~\ref{fig:ALP-summary} in terms of a flavour-universal ALP coupling to quarks ($C^{V,A}_{bs}/f_a = C_{qq}^A/f_a \equiv C_q/f_a$) and its coupling with DM ($C^{A}_{\chi\chi}/f_a$). The ALP mass is set to the value $m_a = 2~{\rm GeV}$ that, according to Ref.~\cite{Fridell:2023ssf}, gives the best fit to $B^+\to K^++{\rm inv}$. The red bands denote the range of $C_q/f_a$ that can account for the Belle~II excess at 1$\sigma$ and $2\sigma$. The contours show the value of $m_\chi$ that fully accounts for the observed relic density, $\Omega_{\chi} h^{2} = 0.12$. Above each line, that is, for larger values of the DM-ALP coupling, the $m_\chi$ needs to be reduced to avoid DM over-abundance. Lyman-$\alpha$ observations set a constraint on how light freeze-in produced DM can be without being in conflict with structure formation, $m_\chi \lesssim 15~\mathrm{keV}$~\cite{Ballesteros:2020adh,DEramo:2020gpr,Decant:2021mhj}, akin to the limits on (thermal) warm DM. As shown in the figure, in combination with the relic density constraint, this sets an upper limit on the DM-ALP coupling, $C^{A}_{\chi\chi}/f_a\lesssim 10^{-5}~{\rm GeV}^{-1}$.
We can also see that, as mentioned above, the natural choice $C^{A}_{\chi\chi}/f_a \approx C_q/f_a \approx (6-7)\times 10^{-9}~{\rm GeV}^{-1}$ is compatible with both the  Belle~II excess and the observed relic abundance if the DM mass is in the MeV range.

As customary, by means of the equation of motion of the singlet field, the derivative interaction of Eq.~\eqref{eq:Lchi} can be rewritten as a pseudoscalar interaction $g_\chi a\,\bar\chi\gamma_5\chi$ with the dimensionless coupling $g_\chi$ given by $g_\chi = \tfrac{m_\chi C^A_{\chi\chi}}{f_a}$. Hence, similar to the other ALP interactions with fermions, the DM-ALP interactions are suppressed by the DM mass. Fig.~\ref{fig:ALP-summary} shows that the range of values of the pseudoscalar coupling consistent with the Belle~II excess and the DM abundance is $g_\chi \sim 10^{-12} - 10^{-10}$. This tiny value guarantees that (i)~DM is never in thermal equilibrium with the visible sector throughout the evolution of the universe, consistently with the premise of the freeze-in mechanism, and (ii)~ALP-DM interactions can not (unfortunately) yield any substantial signal at direct or indirect DM searches, thus trivially evading any existing experimental bound.

The above simplifying assumption of universal ALP-quark couplings can be relaxed without affecting the phenomenology, as long as the ALP is sufficiently long-lived to escape the Belle II detector undetected. If the flavour-conserving ALP-quark couplings are enhanced relative to the flavour-violating one that is needed to explain the Belle~II excess,\footnote{Typically, quark flavour models predict ALPs with off-diagonal couplings suppressed compared to the diagonal ones, see e.g.~\cite{Wilczek:1982rv,Ema:2016ops,Calibbi:2016hwq}.
$|C_{qq}^A|>|C_{sb}^V|$, the lifetime of the ALP will be reduced, since $c\tau_a \propto (|C_{qq}^A|/f_a)^2$, as shown in Eq.~\eqref{eq:ctau}.
However, the ALP decay length is still greater than the size of the Belle~II detector for $f_a/C_{qq}^A \gtrsim 4.6\times 10^8~\mathrm{GeV}$, which is to be considered the lower bound for the mechanism we propose to work. In addition, if $|C_{qq}^A|>|C_{sb}^V|$, the ALP will reach thermal equilibrium earlier, which, however, does not affect freeze-in DM production, because the ALP always reaches thermal equilibrium as we demonstrated above. However, it also reduces the branching ratio for ALP decays to dark matter and thus reduces the DM abundance, which can be compensated by either increasing the DM mass or the ALP-DM coupling.}

\section{Dark photon portal model}
\label{sec:DP}

The dark photon is a well-motivated scenario for a simple dark sector with a $U(1)_X$ gauge symmetry that is spontaneously broken by a dark scalar field $\phi$ of charge $\mathcal{Q}_\phi$.
We briefly review the relevant theory following~Refs.~\cite{Davoudiasl:2012ag,Datta:2022zng,Holdom:1985ag,Babu:1997st}. The Lagrangian of the dark photon model is given by
\begin{align}
    \mathcal{L} \supset \mathcal{L}_{\rm SM} -\frac14 X_{\mu\nu} X^{\mu\nu} + \frac{\varepsilon}{2\cos \theta_W} X_{\mu\nu} B^{\mu\nu} 
    + (D_\mu \phi)^\dagger (D^\mu \phi) - V(\phi) -\kappa H^\dagger H \phi^\dagger \phi \,,
    \label{eq:LagDP}
    \end{align}
where $V(\phi)$ is the scalar potential, $\varepsilon$ is the kinetic mixing parameter between the $U(1)_X$ and the hypercharge field strength tensors, and $D_\mu = \partial_\mu - i g_X \mathcal{Q}_\phi X_\mu$ is the covariant derivative with the gauge coupling $g_X$ and the dark $U(1)_X$ charge $\mathcal{Q}_\phi$. 
After spontaneous symmetry breaking in the dark sector through the vacuum expectation value (VEV) of $\phi$,
\begin{align}
    \phi & = \frac{v_\phi + \varphi+ i a}{\sqrt{2}}\,,
\end{align}
the dark photon acquires a mass $M_X = |\mathcal{Q}_\phi| g_X v_\phi$. As usual, here $\varphi$ is the resulting physical dark Higgs excitation, while $a$ is the would-be Nambu-Goldstone boson providing the massive dark photon with its longitudinal component.
To leading order in the kinetic mixing parameter $\varepsilon$, the kinetic term of Eq.~\eqref{eq:LagDP} is diagonalised by~\cite{Holdom:1985ag}
   \begin{equation}
       B_\mu \to B_\mu + \frac{\varepsilon}{\cos \theta_W} X_\mu\,,
   \end{equation}
and the interactions of the dark photon $X_\mu$ with the SM fermions are thus given by \cite{Davoudiasl:2012ag,Datta:2022zng}
\begin{align}
    \mathcal{L} 
    \supset \left(e\varepsilon J^\text{em}_\mu + \frac{g}{2\cos \theta_W} \varepsilon_Z J^{NC}_\mu\right)X^\mu\,,
\end{align}
in terms of the electromagnetic current $J_\mu^{\rm em}=\sum_f Q_f \bar f \gamma_\mu f$ and the weak neutral current $J_\mu^{\rm NC}=\sum_f (T_{3f}-2Q_f \sin^2\theta_W) \bar f\gamma_\mu f - T_{3f} \bar f\gamma_\mu \gamma_5 f$. 
The quantity $\varepsilon_Z$ parametrises the coupling of the dark photon to the neutral current which receives contributions from the kinetic mixing term and $Z-X$ mass mixing. It is given by 
\begin{equation}\label{eq:epsZ}
\varepsilon_Z \simeq 2 \varepsilon \tan\theta_W \frac{M_X^2}{M_Z^2} \simeq 2\varepsilon \tan\theta_W \left(\frac{m_X^2}{m_Z^2} - \varepsilon \tan\theta_W\right)\;,
\end{equation}
where the physical masses of the $Z$ bosons and the dark photon $X$ are to leading order in $\varepsilon$ given by $m_Z \simeq M_Z$ and $m_X^2 \simeq M_X^2 + \varepsilon \tan\theta_W  M_Z^2$ respectively. For $m_X=2\,\mathrm{GeV}$, the BaBar search for an invisibly decaying dark photon places an upper bound of $\varepsilon< 8.3 \times 10^{-4}$~\cite{BaBar:2017tiz} which translates into an upper bound on $\varepsilon_Z <  3.3\times 10^{-8}$ and thus precludes an explanation of the excess observed by Belle II. The tight relationship between the dark photon coupling to the electromagnetic and neutral current may however be relaxed by a slight extension of the model as it has been discussed in~\cite{Davoudiasl:2012ag}, e.g.~a second isodoublet Higgs $H_2$, which also carries a dark charge additionally contributes to the $Z-X$ mass mixing and thus modifies $\varepsilon_Z$. 
Following~\cite{Davoudiasl:2012ag}, we thus parametrise
\begin{equation}
    \varepsilon_Z = \frac{m_X}{m_Z}\delta\,,
\end{equation}
 where $\delta$ is a model-dependent parameter. As we will see later, an explanation of the Belle II result requires $\varepsilon_Z\simeq 2.3\times 10^{-5}$ and thus $\delta\simeq 1.0\times 10^{-3}$, which is consistent with constraints from atomic parity violation and polarised electron scattering that require $\delta\lesssim 0.01$~\cite{Davoudiasl:2012ag}.\footnote{Within the two Higgs doublet model proposed in~\cite{Davoudiasl:2012ag}, mixing of this size can be easily realised for natural choices of the parameters. For instance, in the limit of small kinetic mixing ($\varepsilon \ll \varepsilon_Z$), $\delta\approx 10^{-3}$ is obtained for a VEV of the second isodoublet Higgs $v_2 \approx 1.5~\mathrm{GeV}$ when $m_X \approx 2~\mathrm{GeV}$, $\alpha_X \approx 10^{-3}$, and $\mathcal{Q}_\phi=\mathcal{Q}_{H_2}=-2$ and thus $v_\phi \approx 10~\mathrm{GeV}$.}

We further introduce two Weyl fermions $\chi_L$ and $\chi_R$ with charge $\mathcal{Q}_\chi=-\mathcal{Q}_\phi/2=1$, which are left-handed and right-handed, respectively. The fermionic Lagrangian reads  
\begin{align}
\mathcal{L} \supset
\bar \chi_L i \slashed{D} \chi_L 
+\bar \chi_R i \slashed{D} \chi_R
-\left[ 
m_D \bar \chi_L \chi_R 
    + \frac{y_L}{2} \overline{\chi_L^c}\chi_L \phi
    + \frac{y_R}{2} \overline{\chi_R^c}\chi_R \phi
    +\mathrm{h.c.}\right] \;.
\label{eq:Lagchi}
\end{align}
After the dark Higgs develops a vacuum expectation value, the Yukawa interactions $y_{L,R}$ result in Majorana mass terms $m_{L,R} = y_{L,R} v_\phi /\sqrt{2}$ for the two Weyl fermions $\chi_{L,R}$.
A similar scenario has been discussed in Ref.~\cite{Garcia:2024uwf}. However, we consider here a different regime of the parameter space where the dark Higgs is lighter than the singlet fermions, so that $\chi_{L,R}$ may annihilate into a pair of dark Higgs particles. 

In general, the fermions $\chi_L$ and $\chi_R^c$ mix with each other through the Dirac mass term $m_D \bar \chi_L \chi_R$, as discussed, e.g.~in Ref.~\cite{Garcia:2024uwf}. We focus on the simplified scenario where the Dirac mass term can be neglected and $\chi_R$ is lighter than the dark photon, which in turn, is lighter than $\chi_L$, i.e.~$0\approx m_D\ll m_R \ll m_X \ll m_L$. This corresponds to the Majorana DM case of Ref.~\cite{Garcia:2024uwf} but with a light dark Higgs. For the following discussion, it is sufficient to consider the lighter of the two fermions. The phenomenology is defined by the Majorana DM candidate $\psi\equiv \chi_R + \chi_R^c$ with mass $m_{\rm DM} \equiv m_R = y_R v_\phi/\sqrt{2}$ and the Lagrangian
\begin{align}
    \mathcal{L} \supset \frac12 \overline{\psi} (i\slashed{\partial} - m_R) \psi  +\frac{g_X}{2} \overline{\psi} \slashed{X} \gamma_5 \psi 
    - \frac{y_R}{2 \sqrt{2}} \overline{\psi} \psi \varphi
    - \frac{y_R}{2 \sqrt{2}} \overline{\psi} i\gamma_5 \psi a
    \;.
\end{align}

\mathversion{bold}
\subsection{Contribution to \texorpdfstring{$B\to K^{(*)} +$ invisible}{B->K(*)+invisible}}
\mathversion{normal}
For light dark photons, the $B$ meson can decay into a kaon $K$ and a dark photon. The lifetime of the dark photon is determined by the kinetic mixing $\varepsilon$ and its coupling to dark fermions, that is, the dark gauge coupling $g_X$. 
The partial decay widths for decays to SM particles are approximately given by
\begin{align}
\Gamma(X\to\ell\bar\ell) &=\frac{\varepsilon^2 \alpha m_X}{3}   \left(1 + \frac{1-4s_W^2}{2 s_W c_W} \frac{\varepsilon_Z}{\varepsilon} + \frac{1-4s_W^2+8 s_W^4}{8 s_W^2 c_W^2} \frac{\varepsilon_Z^2}{\varepsilon^2}\right)\,,
 \\
\Gamma(X\to \nu\bar\nu)  &= \frac{\varepsilon_Z^2 \alpha m_X}{24 s_W^2c_W^2}\,,
\\
\Gamma(X\to \mathrm{hadrons})  &= \Gamma(X\to \mu\bar\mu) \times \mathcal{R}_\mu^h\,,
\end{align}
where $\alpha$ is the fine structure constant, $s_W$ and $c_W$ respectively are the sine and cosine of the weak mixing angle $\theta_W$, $\mathcal{R}^h_\mu \equiv \frac{\sigma(e \bar e \to \text{hadrons})}{\sigma(e\bar e \to \mu\bar\mu)}$ is the $R$-ratio, and we approximated the full expressions in Ref.~\cite{Datta:2022zng} in the limit of small lepton masses, $m_{e,\mu}\ll m_X$, small kinetic mixing $\varepsilon$, and small $\varepsilon_Z$. 
Considering only decays to SM particles, 
we find a proper decay length of 
$c\tau_X =  11 \,(2\times 10^{-5}/\varepsilon)^2\, \mu\mathrm{m}$
for a dark photon with mass $m_X=2~\mathrm{GeV}$ and $\varepsilon=\varepsilon_Z$ and thus $X$  decays promptly inside the detector for the interesting range of $\varepsilon$ (see below). In order to explain the observed excess in $B^+\to K^+ +\mathrm{inv}$, we have to introduce a new invisible decay channel of the dark photon,
because the invisible decay to neutrinos is of the same strength as the decay to charged leptons
and thus cannot explain the observed excess without running in conflict with, e.g., measurements of $B\to K\mu\bar\mu$.
A well-motivated solution is the decay to dark matter. The partial width for decays of the dark photon to dark matter particles is given by  
\begin{align}
    \Gamma(X \to \psi\psi) = \frac{\alpha_X m_X}{6}\left(1-4 \frac{m_{\rm DM}^2}{m_X^2}\right)^{3/2},
    \qquad \text{with}\; \alpha_X=\frac{g_X^2}{4\pi}\;,
\end{align}
which is not suppressed by the kinetic mixing. Numerically, we find that for a dark photon with $m_X=2~\mathrm{GeV}$, $\alpha_X=10^{-3}$, $\varepsilon=\varepsilon_Z= 2\times 10^{-4}$ and a dark matter mass of $m_{\rm DM}=100~\mathrm{MeV}$, the branching ratio to dark matter is effectively 100\%.
As a consequence, the new decay channel to dark matter not only can address the Belle~II excess but also weakens the bounds from visible $B$ decays, such as $B\to K\mu\bar\mu$. Currently, the LHCb measurement of $B\to K\mu\bar\mu$~\cite{LHCb:2014cxe} places the most stringent limit on the kinetic mixing $\varepsilon$ for dark photons with a GeV-scale mass in the absence of additional invisible decay channels.  

As already mentioned, the two-body decay $B \to K X$ can provide a good fit to the observed excess at Belle~II in $B^+\to K^+ +\mathrm{inv}$ for $m_X \approx 2~\mathrm{GeV}$~\cite{Altmannshofer:2023hkn,Fridell:2023ssf}. We thus focus on this scenario in the following. The alternative scenarios with an off-shell dark photon are strongly constrained by searches for $B\to K\mu\bar\mu$, because the invisible decay is suppressed due to three-body decay kinematics.
In the on-shell scenario, the dark photon has a short lifetime and dominantly decays to a pair of DM particles. Hence, the contribution of the dark photon to \mbox{$B\to K^{(*)} +\mathrm{inv}$} is given by the three-body decay width 
\begin{equation}
    \Gamma(B\to K^{(*)} X (\to \psi\psi) ) = \int_{4m_{\rm DM}^2}^{(m_B-m_{K^{(*)}})^2} \frac{dq^2}{2\pi} \frac{[\Gamma(B\to K^{(*)} X) 2 m_X \Gamma(X \to \psi\psi)]_{m_X^2\to q^2}}{(q^2-m_X^2)^2 + m_X^2 \Gamma_X^2}\,,
\end{equation}
which reduces to $\Gamma(B\to K^{(*)} X) \mathrm{BR}(X\to \psi\psi)$ in the narrow width approximation.  

In the SM top-$W$ boson loops generate a monopole coupling of the $Z$ boson to the $b$ and $s$ quarks, as well as dipole couplings of both the photon and the $Z$ boson to $b$ and $s$. The monopole interaction of the photon is forbidden by gauge invariance. Mixing of the dark photon with the photon and the $Z$ boson then results in an interaction of the dark photon with a $b$ and $s$ quark, which induces two-body decay $B\to K^{(*)}X$. The dipole interaction is described by the dimension-6 operator $m_b\, \bar s \sigma^{\mu\nu} P_R b \,F_{\mu\nu}$ and is thus suppressed compared to the renormalizable monopole interaction $\bar s \slashed{X} P_L b$. Hence, up to corrections of order $(m_b/m_W)^2$, the effective vector and axial-vector couplings are equal~\cite{Datta:2022zng}. Neglecting light quarks we find the dark photon effective coupling to $b$ and $s$ is approximately given by
\begin{align}
    g_{bs}^X \simeq \frac{g^3 \varepsilon_Z V_{ts}^* V_{tb}}{64\pi^2 c_W}
    \frac{x_t}{1-x_t}\left( 6-7x_t+x_t^2+(2+3x_t)\ln x_t\right)\,,
    \end{align}
in terms of the weak gauge coupling $g$, the CKM matrix elements $V_{tq}$, and the mass ratio \mbox{$x_t=m_t^2/m_W^2$}. The full expressions in Ref.~\cite{Datta:2022zng} evaluate to 
$|g_{bs}^X|\simeq 1.0 \times 10^{-8} (\varepsilon_Z/2.3\times 10^{-5})$
in good agreement with the above approximate expression. For the analysis in Section~\ref{sec:DPdiscussion}, we use the full numerical expressions. In terms of the above effective coupling, the decay rates take the form 
\begin{align}
\Gamma(B\to KX) & = \frac{|g_{bs}^X|^2 m_B}{64\pi}\frac{\lambda^{3/2}(1,y_K,y_X)}{y_X}|f_+(m_X^2)|^2\,,
\\
\Gamma(B\to K^*X) & = \frac{|g_{bs}^X|^2 m_B}{64\pi}  \lambda^{1/2}(1,y_{K^*},y_X)
\sum_{h=0,\pm} |\tilde H_h|^2\,,
\end{align}
where $y_{K^{(*)}}\equiv m_{K^{(*)}}^2/ m_B^2$ and $y_X \equiv {m_X^2 / m_B^2}$ and the rescaled helicity amplitudes are
\begin{align}
    \tilde H_0 &= - (1 + y^{1/2}_{K^*}) A_1(m_X^2) x_{K^*X} + \frac{2 y^{1/2}_{K^*}y_X^{1/2}}{1+y^{1/2}_{K^*}}A_2(m_X^2) (x_{K^*X}^2-1)\,,\\
    \tilde H_\pm &= (1+y^{1/2}_{K^*}) A_1(m_X^2) +  \frac{2 y_{K^*}^{1/2}y^{1/2}_X}{1+y^{1/2}_{K^*}} V(m_X^2) \sqrt{x_{K^*X}^2-1}
\end{align}
with $x_{K^* X} \equiv (m_B^2-m_{K^*}^2-m_X^2)/(2m_{K^*} m_X)$. The relevant form factors $f_+$, $A_{1,2}$ and $V$ can be obtained from Refs.~\cite{Gubernari:2018wyi,Gubernari:2023puw}. As the decay rates $B\to KX$ and $B\to K^*X$ depend on the same effective coupling, the two decay rates are related to each other with 
\begin{equation}
    \Gamma(B\to K^* X)
    \approx 1.9\;
    \Gamma(B\to KX) 
\label{eq:BtoKs}
\end{equation}
for $m_X=2~\mathrm{GeV}$.
Before presenting the available parameter space in Section~\ref{sec:DPdiscussion}, we discuss the production of DM in the early universe.

\subsection{Dark Matter freeze-out production}
\label{sec:FO}

In the early universe DM is kept in equilibrium with the thermal plasma through interactions mediated by the dark photon. Its abundance is then set by the thermal freeze-out mechanism. The relevant process is the DM annihilation into dark Higgs bosons, $\psi\psi \to\varphi\varphi$, which subsequently decay to SM particles. Following from Eq.~\eqref{eq:Lagchi}, the Majorana DM candidate $\psi$ couples to the dark Higgs boson $\varphi$ via a Yukawa interaction of strength $y_R = \sqrt{2} m_{\rm DM}/v_\phi$ and the annihilation proceeds through a $t$-channel diagram. The annihilation process is $p$-wave suppressed and, hence, the thermally averaged cross section is to leading order in $x^{-1}=T/m_{\rm DM}$ given by 
\begin{align}\label{eq:DMann}
    \left\langle\sigma v_{\text{M\o l}}\right\rangle \simeq 
    \frac{64\pi \alpha_X^2 m_{\rm DM}^2}{m_{X}^4 \, x} \frac{(9-8y+2y^2)\sqrt{1-y} }{(2-y)^4 }\,,  
\end{align}
where $y \equiv m_\varphi^2/m_{\rm DM}^2$. 
The DM relic abundance is inversely proportional to the thermally averaged annihilation cross section
\begin{equation}
    \Omega_\psi h^2 \approx 0.1 \times \left(\frac{0.1\,\mathrm{GeV}}{m_{\rm DM}}\right)^2 \left(\frac{10^{-3}}{\alpha_X}\right)^2
    \qquad \text{for}\qquad y=0.01\;,
\end{equation}
which takes into account that a slightly larger annihilation cross section is required for sub-GeV DM~\cite{Steigman:2012nb}. In Fig.~\ref{fig:dark_photon_flavour} we present the result of a numerical solution of the Boltzmann equation. 
Following~\cite{Kolb:1990vq} we define the deviation from the DM equilibrium yield $\Delta = Y_{\rm DM} - Y^{\rm eq}_{\rm DM}$ and solve
\begin{equation}
   \frac{d\Delta}{d x}  =
   - \frac{d Y_{\rm DM}^{\rm eq}}{d x} 
   - \frac{\lambda}{x^3}  \left(1+\frac13 \frac{d\ln g_*^s}{d\ln T}\right) \Delta (2 Y_{\rm DM}^{\rm eq} + \Delta) 
   \,,
\end{equation}
where $\lambda \equiv \left[ x\langle\sigma v_{\text{M\o l}}\rangle s /H\right]_{x=1}$ is evaluated at $x=1$, $H$ denotes the Hubble rate and $s=2\pi^2 g_*^s(x) T^3/45$ the entropy density and the DM equilibrium yield $Y_{\rm DM}^{\rm eq}$ is defined in Eq.~\eqref{eq:eq-yield}.
Similar to Ref.~\cite{Steigman:2012nb} we read off the final DM relic abundance at $x_f =100\, x_* \simeq 2000$. We use the analytic approximation to the relativistic degrees of freedom presented in App.~A of Ref.~\cite{Wantz:2009it}.

The dark Higgs bosons produced in the annihilation of dark matter particles decay to SM particles. This is enabled by the Higgs portal interaction $\kappa$, which induces mixing between the SM Higgs and the dark Higgs $\varphi$. Beam dump experiments place an upper bound on the resulting scalar mixing angle $\theta$, e.g.~the search for $K^+\to \pi^+ +\mathrm{inv}$ at NA62~\cite{NA62:2021zjw} constrains $\sin\theta\lesssim 1.3\times 10^{-4}$~\cite{Gorbunov:2021ccu,Ferber:2023iso}, while big bang nucleosynthesis (BBN) and other cosmological observations place a lower bound on the mixing angle, see e.g.~Ref.~\cite{Winkler:2018qyg} for a detailed discussion of the parameter space. Here, the most relevant constraint is the lower bound on the dark Higgs mass from the requirement that a thermally produced dark Higgs decays away sufficiently quickly, so that it does not substantially modify BBN. The only available efficient decay channel for MeV-scale dark Higgs bosons is the decay to an electron-positron pair. Hence, the partial decay width takes the form~\cite{Ovchynnikov:2023von}
\begin{align}
    \Gamma(\varphi \to e^+ e^-) = \frac{\sqrt{2}G_F m_e^2 m_\varphi \sin^2\theta}{8\pi}\left(1-\frac{4m_e^2}{m_\varphi^2}\right)^{3/2}\;.
\end{align}
To satisfy BBN constraints, we require the dark Higgs lifetime to be shorter than $\tau_\varphi \lesssim 0.1~\mathrm{s}$. 
Together with the constraint from $K^+\to \pi^++\mathrm{inv}$~\cite{Gorbunov:2021ccu,Ferber:2023iso}, we obtain a lower limit on the dark Higgs mass of $m_\varphi\gtrsim 4~\mathrm{MeV}$. 
We will restrict ourselves to DM masses $m_{\rm DM}\geq 100$ MeV to ensure that freeze-out occurs before BBN. Notice that within this regime astrophysical limits from star cooling and supernova constraints are trivially satisfied~\cite{Raffelt:1990yz}, because $2m_{\rm DM}$ is much larger than the typical temperature of the relevant astrophysical systems.

\subsection{Discussion}
\label{sec:DPdiscussion}

Before presenting the available parameter space, we discuss searches for the Majorana DM candidate $\psi$.
DM direct detection proceeds via $t$-channel exchange of a dark photon that, however, is suppressed because of the axial-vector coupling of the Majorana fermion DM to the dark photon. It is dominated by the vector interaction with the nucleus and thus the nuclear-level matrix element takes the form 
\begin{align}
    \mathcal{M} 
    \simeq - 
    \frac{g_X  b_N}{t-m_{X}^2} (\bar u_{\rm DM} \gamma_\mu \gamma_5 u_{\rm DM}) (\bar u_N \gamma^\mu u_N)\,,
\end{align}
where $b_N  = Z b_p + (A-Z) b_n$ is given in terms of the number of protons $Z$ and nucleons $A$ and the couplings to proton $b_p=2b_u + b_d$ and neutron $b_n=2b_d+b_u$. They can in turn be expressed in terms of the quark couplings
\begin{align}
    b_u & = e\varepsilon\, Q_u + \frac{g}{2c_W} \varepsilon_Z \left(\frac12-2Q_u s_W^2\right) \;,
&
    b_d & = e\varepsilon\, Q_d + \frac{g}{2c_W} \varepsilon_Z \left(-\frac12-2Q_d s_W^2\right) \;,
    \end{align}
    which are dominated by the first term from the kinetic mixing with the dark photon.
The resulting DM-nucleus cross section is velocity suppressed and given by 
\begin{equation}
    \sigma_N = \frac{2\alpha_X b_N^2 v_{\rm DM}^2}{m_{X}^2}\frac{x_N^2 x^2_{\rm DM} (x_N^2+2 x_N x_{\rm DM} +3 x_{\rm DM}^2)}{(x_N+x_{\rm DM})^4}\,,
\end{equation}
with $x_{N,\rm DM} = m_{N,\rm DM}/m_{X}$ and the DM velocity $v_{\rm DM}$.
For DM scattering off a Xenon nucleus, $x_N\gg x_{\rm DM}$, and the cross section is approximated by  
\begin{equation}
    \sigma_N 
    \approx 6.7 \times 10^{-40} 
    \left(\frac{\varepsilon}{8.3\times 10^{-4}}\right)^2\frac{\alpha_X}{10^{-3}} 
    \left(\frac{m_{\rm DM}}{\mathrm{GeV}}\right)^2 \left(\frac{v_{\rm DM}}{235 \,\mathrm{km}/\mathrm{s}}\right)^2 \mathrm{cm}^2\;.
\end{equation}
Currently, the strongest limit for DM in this mass range is set by the Panda-X experiment~\cite{PandaX:2023xgl} using the Migdal effect, which sets the limit 
$\sigma_N\lesssim 1.8\times 10^{-39}\,\mathrm{cm}^2$ at $m_{\rm DM}=1~\mathrm{GeV}$, which is rapidly weakened for lighter DM masses. Hence, DM direct detection searches are currently not sensitive to the preferred parameter space.

DM scattering off electrons does not pose any constraint. The cross section for scattering off a free electron can be obtained from the DM nucleus scattering cross section written above by replacing $b_N \to -\varepsilon e + \varepsilon_Z g (-1+4 s_W^2)/4c_W$. For DM much heavier than the electron, i.e.~in the limit $x_N\ll x_{\rm DM}$, the cross section takes the form
\begin{equation}
    \sigma_e 
        \approx 3.7
        \times 10^{-52} 
    \left(\frac{\varepsilon}{8.3\times 10^{-4}}\right)^2\frac{\alpha_X}{10^{-3}} 
     \left(\frac{m_{\rm DM}}{\mathrm{GeV}}\right)^2 
     \left(\frac{v_{\rm DM}}{235\, \mathrm{km}/\mathrm{s}}\right)^2 
     \mathrm{cm}^2\;.
\end{equation}

Finally, we find that there are no constraints from indirect detection, because the dominant DM annihilation into a pair of dark Higgs bosons is $p$-wave suppressed, as shown above, and no conflict with bounds on DM self-interactions, despite the value of the dark gauge coupling consistent with the observed relic abundance being rather large, $\alpha_X\approx 10^{-3}$, because of the relatively sizeable mass of the mediator of the interaction, that is, the dark photon. 
In the non-relativistic limit, the DM self-interaction cross section is 
\begin{equation}
\begin{aligned}
    \frac{\sigma_{\rm SI}}{m_{\rm DM}} 
    &= \frac{16\pi \alpha_X^2 v_{\rm DM}^2}{m_{X}^3} \frac{x_{\rm DM} (3-20x_{\rm DM}^2+40x_{\rm DM}^4)}{(1-4x_{\rm DM}^2)^2} 
    \\ &
    \approx  3.5\times 10^{-15}  \left(\frac{\alpha_X}{10^{-3}}\right)^2 \left(\frac{v_{\rm DM}}{3000 \,\mathrm{km}/\mathrm{s}}\right)^2  \frac{x_{\rm DM} (3-20x_{\rm DM}^2+40x_{\rm DM}^4)}{(1-4x_{\rm DM}^2)^2} \frac{\mathrm{cm}^2}{\rm g} \;,
\end{aligned}
    \end{equation} 
    for $m_X=300$ MeV and a relative DM velocity of 3000 km/s~\cite{Cirelli:2024ssz}. This is far below the galaxy cluster bound of $\sigma/m_{\rm DM} \lesssim 1.25\, \mathrm{cm}^2/\mathrm{g}$~\cite{Randall:2008ppe}.

\begin{figure}[bt!]
    \centering
    \includegraphics[width=0.48\linewidth]{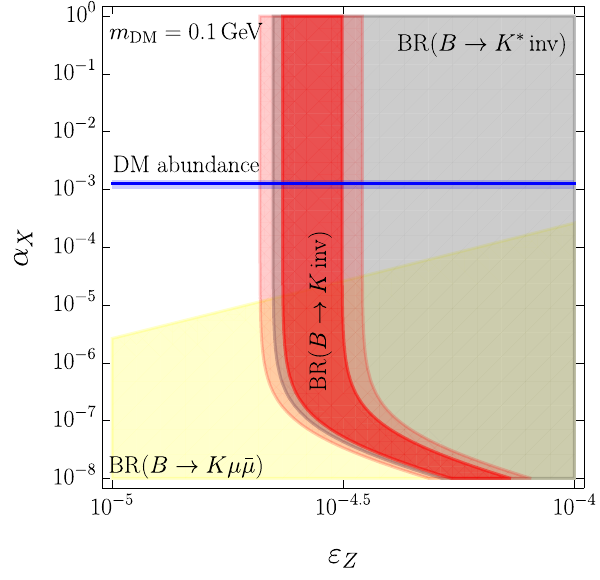}%
    \hfill
    \includegraphics[width=0.48\linewidth]{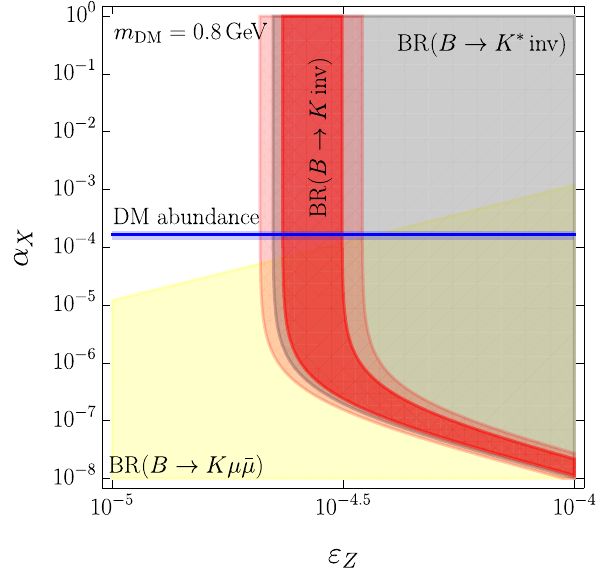}
    \caption{Dark photon parameter space for $m_{\rm DM}=0.1~\mathrm{GeV}$ (\emph{left}) and $m_{\rm DM}=0.8~\mathrm{GeV}$ (\emph{right}).  The yellow and gray regions are excluded by searches for $B^+\to K^+\mu\bar\mu$~\cite{LHCb:2014cxe,Datta:2022zng} and $B^0\to K^{*0} +\mathrm{inv}$~\cite{Belle:2017oht}, respectively. The kinetic mixing has been fixed at the maximum allowed value of $\varepsilon =8.3\times 10^{-4}$~\cite{BaBar:2017tiz}. Smaller values weaken the LHCb constraint. 
    The Belle~II $B^+\to K^++\mathrm{inv}$ excess is accounted for at the $1\sigma$ and $2\sigma$ level in the red regions. The blue bands show the region of parameter space where the DM relic abundance is explained by the Majorana singlet $\psi$. The solid blue line corresponds to a dark Higgs with mass $m_\varphi=4~\mathrm{MeV}$ and the blue shaded region displays the range $m_{\varphi} \in [4\,\mathrm{MeV},0.99\,m_{\rm DM}]$. 
    }
    \label{fig:dark_photon_flavour}
\end{figure}

The results are summarized in Fig.~\ref{fig:dark_photon_flavour}. As discussed above, the decay rate of $B\to K^{(*)} +\mathrm{inv}$ is governed by $\varepsilon_Z$ which is induced by $Z-X$ mass mixing, while the visible decay $B\to K \mu\bar\mu$ depends on both $\varepsilon$ and $\varepsilon_Z$ through the leptonic decay width of $X$. In the figure, the kinetic mixing has been fixed at the maximum allowed value $\varepsilon=8.3\times 10^{-5}$~\cite{BaBar:2017tiz}. Smaller values suppress the branching ratio of $B^+\to K^+\mu\bar\mu$ and therefore weaken the constraint obtained by LHCb~\cite{LHCb:2014cxe,Datta:2022zng}. For $m_X=2~\mathrm{GeV}$,
the dark photon can explain the excess in $B^+\to K^+ +\mathrm{inv}$ at $1\sigma$ ($2\sigma$) in the region shown in (light) red,
which was obtained using $[0.5,0.9]\times 10^{-5}$ ($[0.4,1.1]\times 10^{-5}$) for the new physics contribution~\cite{Fridell:2023ssf}. The gray and yellow regions are excluded by searches for $B\to K^* + \mathrm{inv}$ that place the upper limit $\mathrm{BR}(B^0\to K^{*0} +\mathrm{inv})<1.8\times 10^{-5}$~\cite{Belle:2017oht} and by the LHCb search for $B^+\to K^+\mu\bar\mu$~\cite{LHCb:2014cxe} respectively. Combining the latter results with the SM predictions quoted in Tab.~1 of Ref.~\cite{Datta:2022zng}, we impose the upper limit of $\mathrm{BR}(B^+\to K^+ X (\to \mu \bar\mu))\leq 3.6\times 10^{-9}$ for $m_X = 2~\mathrm{GeV}$. As we can see, our dark-photon scenario is unable to explain the Belle II excess at $1\sigma$ because of the tight constraint from $B\to K^* + \mathrm{inv}$, but it is consistent with the observation at $2\sigma$.

Finally, let us note that the observed cosmic DM abundance is governed by the dark fine structure constant $\alpha_X$ and the DM mass $m_{\rm DM}$. For $\alpha_X\lesssim 10^{-4}$,
the DM abundance overcloses the universe, because the dark Higgs VEV is much larger than the DM mass. The correct DM relic abundance $\Omega_\psi h^2 = 0.12$ 
is obtained within the narrow blue bands for dark Higgs masses $m_\varphi \in [4 \,\mathrm{MeV},0.99\,m_{\rm DM}]$. Tuning the dark Higgs mass further $m_\varphi \simeq m_{\rm DM}$ allows larger values of $\alpha_X$. The solid blue line assumes a dark Higgs mass of $m_\varphi=4~\mathrm{MeV}$, which is the lowest possible value consistent with BBN and direct searches as discussed in Section~\ref{sec:FO} above.

\section{Conclusions}
\label{sec:Conclusions}
We presented two simple models that may explain the excess in $B^+\to K^+ +\mathrm{inv}$ observed by Belle~II in terms of $B$ decaying to a bosonic mediator which subsequently decays to fermionic dark matter able to account for 100\% of the observed relic abundance. Both scenarios address the excess in terms of the two-body decay into a $\approx 2~\mathrm{GeV}$ boson, which was shown to provide a good fit to the data~\cite{Fridell:2023ssf}. The ALP model is able to account for the observed excess in $B^+\to K^++\mathrm{inv}$ below $1\sigma$, while the dark-photon scenario is more tightly constrained by $B \to K^* +\mathrm{inv}$ and thus can only account for the excess at the $2\sigma$ level.

The models, however, differ in the lifetime of the mediator, and the production of dark matter in the early universe. While the ALP is long-lived and escapes the detector, the dark photon decays promptly into dark matter. Furthermore, the vastly different interaction strength of the mediator-DM coupling governs the DM production in the early universe.  

In the dark photon model, the Belle~II excess requires a sizeable kinetic mixing that allows the dark matter to fully thermalise. Hence, DM production is based on the freeze-out mechanism and the relic abundance is set by the cross section of the DM annihilation into a pair of dark Higgs particles that subsequently decay to light SM particles.

Conversely, in the ALP scenario, the DM candidate is feebly coupled and produced through the freeze-in mechanism from ALP decays. The ALP itself is produced through $b$-quark decays, $b\to s a$ and $2-2$ scattering processes like $b g \to ba$ and $b\bar b \to g a$, and reaches thermal equilibrium for temperatures close to the $b$-quark mass. Hence, the DM abundance is largely insensitive to the ALP couplings to SM fermions as long as they are sufficiently large to guarantee the thermalisation of the ALP. 
We demonstrated that universal fermionic ALP couplings with $C_q/f_a =C_{\chi\chi}^A/f_a \simeq (6-7)\times 10^{-9}\, \mathrm{GeV}^{-1}$ are able to explain both the dark matter relic abundance and the observed excess in $B^+\to K^+ +\mathrm{inv}$. 

Current dark matter direct detection experiments are unable to probe the available parameter space. Future DM direct detection experiments may be able to probe the dark-photon scenario using the Migdal effect. The current exclusion limit of the Panda-X experiment is close to the preferred parameter space for DM masses close to $1$ GeV. 

The recent analysis of $B^+\to K^+ +\mathrm{inv}$ events by Belle II~\cite{Belle-II:2023esi} was only the first measurement of this kind. Belle II is also expected to measure other rare $B$ meson decays, such as $B\to K^{*}+\mathrm{inv}$ reaching an ultimate sensitivity down to $\mathcal{O}(10\%)$ of the SM prediction~\cite{Belle-II:2022cgf}. A future measurement of $B\to K^* + \mathrm{inv}$ will directly probe our dark-photon model, since the decay rate is directly correlated to $B\to K +\mathrm{inv}$, see Eq.~\eqref{eq:BtoKs}. There is no such tight correlation in the ALP scenario that we considered. However, an ultraviolet (UV) completion of the model may introduce a connection among the chiral ALP couplings to fermions thus increasing the predictivity. Finally, we note that the light dark Higgs boson in the dark photon model might contribute to $K^+\to \pi^+ +\mathrm{inv}$ and partially explain the mild excess recently observed by NA62~\cite{NA62:2024pjp} via the two-body decay $K^+\to \pi^+ \varphi$. We leave the discussion of these interesting extensions of our models to future work.

\section*{Acknowledgements}
We would like to thank Thomas Browder, Slavomira Stefkova, Bruce Yabsley and Robert Ziegler for useful discussions. We are also grateful to Felix Kahlh\"ofer for pointing out a mistake in the definition of $\varepsilon_Z$ in the first arXiv version of the paper, Susanne Westhoff for pointing out the contribution of undetected visible ALP decays to $B\to K+\mathrm{inv}$, and Jongkuk Kim for pointing out the BaBar search for invisibly decaying dark photons $X$ in $e^+e^- \to \gamma X$.
LC and TL acknowledge financial support from the National Natural Science Foundation of China (NSFC) under the grant No.~12035008. 
TL is also supported by the NSFC under the grants No.~12375096 and 11975129. 
MS acknowledges support from the Australian Research Council Discovery Project DP200101470.

\bibliography{refs}

\end{document}